\documentclass[11pt,reqno,sumlimits,intlimits]{amsart}
\usepackage{amsmath}
\usepackage{amssymb}
\usepackage{amsthm,amsxtra}
\newtheorem{theorem}{Theorem}
\newtheorem{lemma}[theorem]{Lemma}
\newtheorem{proposition}[theorem]{Proposition}

\newtheorem{corollary}[theorem]{Corollary}

\theoremstyle{remark}

\newtheorem*{acknow}{Acknowledgments}

\newcommand{\RRe}{\ensuremath{\mathrm{Re}}}
\newcommand{\IIm}{\ensuremath{\mathrm{Im}}}
\newcommand{\im}{\ensuremath{\mathrm{i}}}
\newcommand{\row}{\ensuremath{\mathrm{Row}}}
\newcommand{\TT}{\ensuremath{\mathrm{t}}}
\newcommand{\tr}{\ensuremath{\mathrm{tr}}}

\newcommand{\XX}{\ensuremath{\mathbf{X}}}

\newcommand{\II}{\ensuremath{\boldsymbol{1}}}
\newcommand{\Res}[1]{\ensuremath{\underset{#1}{\mathrm{Res}}}}

\begin{document}

\title{A note on biorthogonal ensembles}

\author{Patrick  Desrosiers}
\address{Department of Mathematics and Statistics, University of
Melbourne, Parkville, Victoria 3010, Australia.}
\email{P.Desrosiers@ms.unimelb.edu.au}
\author{Peter J. Forrester}
\address{Department of Mathematics and Statistics, University of
Melbourne, Parkville, Victoria 3010, Australia.}
\email{P.Forrester@ms.unimelb.edu.au}

\date{August 2006}

\begin{abstract}

We consider ensembles of random matrices, known as biorthogonal
ensembles, whose eigenvalue probability density function can be
written as a product of two determinants.  These systems are closely
related to multiple orthogonal functions.  It is known that the
eigenvalue correlation functions of such ensembles can be written as
a determinant of a kernel function.  We show that the kernel is
itself an average of a single ratio of characteristic polynomials.
In the same vein, we prove that the type I multiple polynomials can
be expressed as an average of the inverse of a characteristic
polynomial.  We finally introduce a new biorthogonal matrix
ensemble, namely the chiral unitary perturbed by a source term.
\end{abstract}

\maketitle

\tableofcontents

\section{Introduction}

Suppose that we have a set of $N$ real random variables
$\{x_1,\ldots,x_N\}$ such that their probability density function
(p.d.f.) is given by
\begin{equation}\label{biorthopdf}
p_N(x_1,\ldots,x_N)=\frac{1}{Z_N}\det
\left[\eta_i(x_j)\right]_{i,j=1,\ldots,N}\det
\left[\xi_i(x_j)\right]_{i,j=1,\ldots,N},
\end{equation}
where $Z_N$ is the normalization constant.  We require all variables
to lie on the same interval $I\subseteq \mathbb{R}$. Consider the
$n$-point correlation functions 
\begin{equation}
\rho_{n,N}(x_1,\ldots,x_n):=\frac{N!}{(N-n)!}\frac{1}{Z_N}\int_I
dx_{n+1}\ldots\int_I dx_N\, p_N(x_1,\ldots,x_n,x_{n+1}\ldots,x_N).
\end{equation}
Now assume that the matrix $\mathbf{g}$, with elements
$g_{i,j}:=\int_I dx\, \eta_i(x)\xi_j(x)$, is not singular.  Then,
one can show that the $n$-point correlation functions can be written
as the determinant of an $n\times n$ matrix:
\begin{equation}\label{corrK}
\rho_{n,N}(x_1,\ldots,x_n)=\det\left[K_N(x_i,x_j)\right]_{i,j=1,\ldots,n},
\end{equation}
where the function $K_N$, called the kernel, is given by
\begin{equation}\label{Kdouble}
K_N(x,y)=\sum_{i,j=1}^N\eta_i(x)c_{i,j}\xi_j(y),\qquad
\sum_{k=1}^Ng_{i,k}c_{j,k}:=\delta_{i,j}.
\end{equation}
Subject to a minor technical constraint on $\mathbf{g}$ (see \cite{Borodin}),
it is possible to construct
functions $\zeta_i\in\mathrm{Span}(\xi_1,\ldots,\xi_N)$ and
$\psi_j\in\mathrm{Span}(\eta_1,\ldots,\eta_N)$ which are
biorthogonal; that is, \begin{equation} \int_I
dx\,\psi_i(x)\zeta_j(x)=\delta_{i,j}
\end{equation}
As a consequence, we can put the kernel in a single sum form:
$K_N(x,y)=\sum_{i=1}^N\psi_i(x)\zeta_i(y)$.

Borodin \cite{Borodin} has used the expression ``biorthogonal
ensembles" for describing systems whose p.d.f.\ can be written as in
Eq.\ \eqref{biorthopdf}. They have been first studied by Muttalib
\cite{Muttalib} and Frahm \cite{Frahm} in relation to the quantum
transport theory of disordered wires \cite{Beenakker}. They can also
be considered as determinantal point processes \cite{Johansson}.

Random Matrix Theory \cite{Deift,ForresterBook,Metha} provides many
instances of such biorthogonal structures. First, choose
$\eta_j=x^{j-1}$ and $\xi_j(x)=e^{-V(x)}x^{j-1}$.  Then
$p_N(x_1,\ldots,x_N)$ corresponds to the eigenvalue density in a
unitary invariant ensemble of $N\times N$ complex Hermitian
matrices, defined through the p.d.f.
\begin{equation}\label{unitarypdf}
P_N(\mathbf{X})\propto e^{-\mathrm{Tr}V(\mathbf{X})}.
\end{equation}
In that case, the simplest one, the system is described by
orthonormal polynomials $\{p_{i}:i=0,\ldots,N-1\}$ with respect to
the weight $w(x)=e^{-V(x)}$; explicitly, $\psi_i(x)=p_{i-1}(x)$ and
$\zeta_i(x)=w(x)p_{i-1}(x)$.

Second, suppose that the we break the unitary invariance of
\eqref{unitarypdf} by an external source \cite{BH1,BH2},
\begin{equation}\label{pertupdf}
P_N(\mathbf{X}|\mathbf{A})\propto
e^{-\mathrm{Tr}V(\mathbf{X})+\mathrm{Tr}\mathbf{A}\mathbf{X}},
\end{equation}
where $\mathbf{A}$ is a non-random $N\times N$ Hermitian matrix.
When the eigenvalues $\{a_1,\ldots,a_N\}$ of $\mathbf{A}$ are all
distinct, it is possible to show that the p.d.f.\ for the eigenvalue
of $\mathbf{X}$ is of the form \eqref{biorthopdf}, with
$\eta_i(x)=x^{i-1}$ and $\xi_i(x)=e^{-V(x)+a_i x}$.  It  has been
proved by Zinn-Justin \cite{Zinn1,Zinn2} that the $n$-point
correlation functions comply with \eqref{corrK}.  In the general
case where some of the parameters are equal, Bleher and Kuijlaars 
\cite{BKmulti} have shown that the models defined by
\eqref{unitarypdf} naturally lead to multiple orthogonal polynomials
(see for instance \cite{Aptekarev,VAC}). In particular,  they have
proved that the (monic) multiple polynomial of type II and having
degree $N$, $P(x)$ say, is simply described as the expectation value
of the characteristic polynomial,
\begin{equation}
P(x)=\langle\det(x\II-\mathbf{X})\rangle,
\end{equation}
where the average is taken with respect to the p.d.f.\ \eqref{pertupdf}.

In this paper, we show that the multiple polynomials of type I, here
denoted by $Q(x)$, can also be seen as averages over perturbed
matrix ensembles,
\begin{equation}\label{Q}
Q(x)=\Res{z=x}\left\langle{\det(z\II-\mathbf{X})}^{-1}\right\rangle
\quad\mbox{for}\quad z\in\mathbb{C}\setminus\mathbb{R}.
\end{equation}
In the previous equation, the residue is defined through
\begin{equation}\label{res}
f(x)=:\underset{z=x}{\mathrm{Res}}\int_I dt\, \frac{f(t)}{z-t} .
\end{equation}
We also obtain a similar expression for the kernel,
\begin{equation}
K_N(x,y)=\frac{1}{x-y}\,\underset{z=y}{\mathrm{Res}}\,\left\langle\frac{\det(x\II-\mathbf{X})}{\det(z\II-\mathbf{X})}\right\rangle.
\end{equation}
This expression was first proposed in \cite{Guhr} for Gaussian
ensembles (i.e., $V(\mathbf{X})=\mathbf{X}^2$), and for general
unitary invariant potentials \eqref{unitarypdf} in \cite{BS}.

We also give a matrix model that possesses a new biorthogonal
structure: the perturbed chiral Gaussian unitary ensemble (chGUE). The chiral
or Laguerre ensemble plays a fundamental role in the low energy
limit of QDC \cite{Verbaarschot}.  It also appears in multivariate
statistics; more specifically, a chiral ensemble is equivalent to
the matrix variate Wishart distribution. The presence of a source
term in the matrix model describes a non-null sample covariance
matrix \cite{Johnstone}.  For more information on this relation, see
\cite{Baik2004,DF0406}. For the parameter value $\alpha = 1/2$ the perturbed
chGUE gives the p.d.f.~for non-intersecting Brownian paths near a wall
\cite{Katori}, and similarly for $\alpha$ a non-negative integer it 
corresponds to non-colliding systems of $2(\alpha + 1)$-dimensional squared
Bessel processes \cite{Konig}.
The biorthogonal functions of the perturbed
chGUE are related to the modified Bessel functions of the first kind.
In a special case, these functions previously appeared in papers by
Coussement and Van Assche \cite{CVA1,CVA2}.

\section{Kernel and ratio of characteristic polynomials }

For any ensemble composed of matrices having real eigenvalues, it is
well known (see e.g.~\cite{BS}) the correlations functions can be generated by averaging
ratios of characteristic polynomials:
\begin{equation}
\rho_{n,N}(x_1,\ldots,x_n)=\Res{z_1=x_1} \ldots\Res{z_n=x_n} \left[\frac{\partial^n}{\partial y_1\cdots\partial y_n}\left\langle\prod_{i=1}^n\frac{\det(y_i\II-\XX)}{\det(z_i\II-\XX)}\right\rangle\right]_{y_i=z_i}.
\end{equation}
This formula can be proved by using
\begin{equation}
\det(y\II-\XX)^{-1}\frac{\partial }{\partial y} \det(y\II-\XX)= \tr
\frac{1}{y\II-\XX},
\end{equation}
and by expressing the matrix average as an average over the
eigenvalues.

In the physics literature, the residue operation is often replaced
by the use of Green's functions and density operators. This can be
understood as follows. The $n$-point correlation function is the
expectation value of a product of $n$ density operators; that is,
\begin{equation}
\rho_{n,N}(x_1,\ldots,x_n)=\left\langle\hat\rho(x_1)\cdots\hat\rho(x_n)\right\rangle,
\end{equation}
where
\begin{equation}
 \hat{\rho}:=\tr
\,\delta(x\II-\XX)
\end{equation}
and it is assumed that the points are not coincident.  For our
purposes, the Dirac delta function has to be defined as
\begin{equation}
\delta(x)=\frac{1}{\pi}\frac{\epsilon}{x^2+\epsilon^2}=\frac{1}{\pi}\IIm\,
\frac{1}{x-\im\epsilon},\qquad \epsilon\rightarrow0^+.
\end{equation}
But the advanced Green function is given by
\begin{equation} \hat
G^-(x):=\tr \frac{1}{(x-\im\epsilon)\II-\XX}.
 \end{equation}
  Hence
\begin{equation} \hat\rho(x)=\frac{1}{\pi}\IIm\, \hat G^-(x).
\end{equation}
It is worth mentioning that the previous formalism allows us to
rewrite Eq.\ \eqref{res} as
\begin{equation}
\underset{z=x}{\mathrm{Res}}\int_I dt\,
\frac{f(t)}{z-t}=\frac{1}{\pi}\IIm \int_I dt\,
\frac{f(t)}{x-\im\epsilon-t},\qquad \epsilon\rightarrow 0^+.
\end{equation}
The imaginary part can be taken in two ways: 1) forming the Dirac
function $\delta(x)$ inside the integrand, then integrating; 2)
deforming the contour of integration, in order to remove the
imaginary part from the integrand, then subtracting the imaginary
part of the whole integral, which is equivalent to closing the
contour around the single pole $x$.

\begin{proposition}Consider a matrix model with an eigenvalue p.d.f.~given by Eq.\ \eqref{biorthopdf}.
Let $\eta_i(x)=x^{i-1}$ and $\xi_i(x)$ such that $g_{i,j}:=\int_I
dx\, \eta_i(x)\xi_j(x)$ defines a non-singular matrix for all
$i,j=1,\ldots N$. Then
\begin{equation}\label{Kaverage}
K_N(x,y)=\frac{1}{x-y}\,\Res{z=y}\,\left\langle\frac{\det(x\II-\mathbf{X})}{\det(z\II-\mathbf{X})}\right\rangle,
\end{equation}
where $z$ is a complex number with a non-null imaginary part.
\end{proposition}
\begin{proof}
We want to prove that the previous expression for $K_N(x,y)$ is
equivalent to Eq.\ \eqref{Kdouble}.  Let us denote the r.h.s.\ of
\eqref{Kaverage} as $(x-y)^{-1}Z_N^{-1}L_N(x,y)$.   From
\eqref{biorthopdf} we have
\begin{equation*}
L_N(x,y)= \Res{z=y}\int_I dx_1\cdots\int_I dx_N\, \det
\left[\xi_i(x_j)\right]_{i,j=1}^N\det
\left[x_j^{i-1}\right]_{i,j=1}^N\prod_{i=1}^N\frac{x-x_i}{z-x_i}\,.
\end{equation*}
By symmetry of the integrand, this can be simplified
\begin{equation}\label{L}
L_N(x,y)=  N!\, \Res{z=y}\int_I dx_1\, \xi_1(x_1)\cdots\int_I dx_N\,
\xi_N(x_N) \det
\left[x_j^{i-1}\right]_{i,j=1}^N\prod_{i=1}^N\frac{x-x_i}{z-x_i}\,.
\end{equation}
But one proves by induction that
\begin{equation}\label{formula1}
\prod_{i=1}^N\frac{1}{z-x_i}=\sum_{i=1}^N\frac{1}{(z-x_i)}
\prod_{\substack{j=1\\j\neq i}}^N\frac{1}{x_i-x_j}. \end{equation}
Moreover,
$\det\left[x_j^{i-1}\right]_{i,j=1}^N=\Delta(x_1,\ldots,x_N)=\prod_{1\leq
i< j\leq N}(x_j-x_i)$ is the Vandermonde determinant.  From this we
deduce
\begin{equation}\label{formula2}
\prod_{\substack{j=1\\j\neq
i}}^N\frac{1}{x_i-x_j}\det\left[x_j^{i-1}\right]_{i,j=1}^N
=(-1)^{N-i}\Delta^{(i)}(x_1,\ldots,x_N),
\end{equation}
 where $\Delta^{(i)}(x_1,\ldots,x_N)=\Delta(x_1,\ldots,x_{i-1},x_{i+1},\ldots,x_N)$.
By substituting formulae \eqref{formula1} and \eqref{formula2} into
\eqref{L}, we get
\begin{multline*}
L_N(x,y)=  N!\, \sum_{i=1}^N(-1)^{N-i}\Res{z=y} \int_I dx_1\,
\xi_1(x_1)\cdots\int_I dx_N\, \xi_N(x_N) \frac{x-x_i}{z-x_i}\\
\times \prod_{\substack{j=1\\j\neq
i}}^N(x-x_j)\det\left[x_j^{k-1}\right]_{\substack{j=1,\ldots,N\\k=1,\ldots,N-1\\
j\neq i}}\,.
\end{multline*}The two factors on the last line can be replaced by the Vandermonde
\[\Delta(x_1,\ldots,x_{i-1},x_{i+1},\ldots,x_N,x)\]
Taking the residue then gives
\begin{equation*}
L_N(x,y)= (x-y) N!\, \sum_{i=1}^N(-1)^{N-i}\xi_i(y)
\left(\prod_{\substack{j=1}}^N\int_I dx_j\xi_j(x_j)\, \det\left[
\begin{array}{c}
[x_j^{k-1}]_{\substack{j=1,\ldots,N\\k=1,\ldots,N}} \cr
[x^{k-1}]_{k=1,\ldots,N}\end{array} \right]\right)_{j\neq i}.
\end{equation*}
Recalling $\eta_i(x)=x^{i-1}$, $g_{i,j}=\int_I dx\,
\eta_i(x)\xi_j(x)$, and integrating the determinant row by row, we
find
\begin{equation*}
L_N(x,y)= (x-y) N!\, \sum_{i=1}^N\xi_i(y) \, \det\left[
\big[g_{j,k}\big]_{\substack{j=1,\ldots,N\\k=1,\ldots,i-1}}
\big[\eta_j(x)\big]_{j=1,\ldots,N}\big[g_{j,k}\big]_{\substack{j=1,\ldots,N\\k=i+1,\ldots,N}}
\right] .
\end{equation*}
We now return to $K_N(x,y)=(x-y)^{-1}Z_N^{-1}L_N(x,y)$.  From
the p.d.f.\ \eqref{biorthopdf}, we have
\begin{equation}\label{ZN}
Z_N=N!\det\mathbf{g}, \qquad
\mathbf{g}^\TT=\mathbf{c}^{-1}.\end{equation} This leads to
\[K_N(x,y)=\sum_{i=1}^N\xi_i(y)\,\det \mathbf{K}^{(i)},\]
where
\[\mathbf{K}^{(i)}=\left[\begin{array}{lll}\II_{(i-1)\times
(i-1)}&\big[\sum_{k=1}^N\eta_{k}(x)c_{k,j}\big]_{j=1,\ldots,i-1}&\mathbf{0}_{(i-1)\times(N-i)}\\
\mathbf{0}_{(N-i+1)\times(i-1)}&\big[\sum_{k=1}^N\eta_{k}(x)c_{k,j}\big]_{j=i,\ldots,N}&\II_{(N-i+1)\times
(N-i)}\end{array}\right]\]Therefore
$K_N(x,y)=\sum_{i,k=1}^N\xi_i(y)\eta_k(x) c_{k,i}$, as desired.
\end{proof}

Before going to the next section, let us show that Eq.\
\eqref{Kaverage} readily furnishes the Christoffel-Darboux formula
for orthogonal polynomials.  We define the orthogonal ensemble by
$\eta_{i}(x)=x^{i-1}$ and $\xi_{i}(x)=x^{i-1}w(x)$, where $w(x)>0$
is the unnormalized weight function. Let
$p_i(x)=x^i+c_1x^{i-1}+\ldots$ denote the monic orthogonal
polynomial with 
\[ h_n\delta_{n,m}=\int_Idx\,w(x)p_n(x)p_m(x).\]
We find $\det[x_i^{j-1}]=\det[p_{j-1}(x_i)]$ and
$Z_N=N!\prod_{i=0}^{N-1}h_n$. On the one hand, by proceeding as in
the proof Proposition 1, we find
\begin{multline}\label{CD1}
   K_N(x,y)=\frac{1}{x-y}\frac{1}{\prod_{i=0}^{N-1}h_n}\Res{z=y}\sum_{i=1}^N\int_Idx_1\,w(x_1)p_0(x_1)\ldots\int_I dx_N
   w(x_N)p_{N-1}(x_{N-1})\\
   \times\frac{x-x_i}{z-x_i}\det\left[
   \big[p_{j-1}(x_k)\big]_{\substack{j=1,\ldots,N\\k=1,\ldots i-1}}\big[p_{j-1}(x)\big]_{j=1,\ldots,N}\big[p_{j-1}(x_k)\big]_{\substack{j=1,\ldots,N\\k=i+1,\ldots N}}\right]
.\end{multline} By virtue of the orthogonality of the $p_n$'s, the
latter equation is equivalent to
\begin{equation}\label{CD1}K_N(x,y)=w(y)\sum_{n=0}^{N-1}\frac{1}{h_n}p_n(y)p_n(x).
\end{equation}
On the other hand, making use of
\begin{equation}\label{formula3}
\frac{1}{\prod_{i=1}^N(z-x_i)}\det\left[\xi_j(x_i)\right]_{i,j=1}^N
=\prod_{i=1}^Nw(x_i)\det\left[\begin{array}{cccc}
  p_0(x_1) & \ldots & p_{N-2}(x_1)&\frac{1}{z-x_1}  \\
  p_0(x_2) & \ldots & p_{N-2}(x_2)& \frac{1}{z-x_2} \\
  \vdots & \vdots  & \vdots  & \vdots \\
  p_{0}(x_N)&\ldots & p_{N-2}(x_N)&\frac{1}{z-x_N}
 \end{array}\right],
\end{equation}
we get
\begin{multline}
(x-y)\left(\prod_{n=0}^{N-1}h_n\right)
K_N(x,y)=\\
\Res{z=y}\int_Idx_1\,\frac{w(x_1)}{z-x_1}\int_Idx_2\,w(x_2)p_0(x_2)\ldots\\
\int_I
dx_N
   w(x_N)p_{N-2}(x_{N})
   \det\left[\begin{array}{ccc}
  p_0(x_1) & \ldots & p_{N}(x_1) \\
  \vdots & \vdots  & \vdots   \\
 p_{0}(x_N)&\ldots&p_{N}(x_N)\\
  p_{0}(x)&\ldots&p_{N}(x)
 \end{array}\right].
\end{multline}
We finally integrate the determinant row by row and arrive at
\[(x-y)\left(\prod_{n=0}^{N-1}h_n\right)
K_N(x,y)=w(y)\det\left[\begin{array}{cccccc}
  h_0&0&\ldots&0&0&0\\
  0&h_1&\ddots&0 &0 &0   \\
  \vdots&\ddots&\ddots&\vdots&\vdots&\vdots\\
  0&\ldots&\ldots&h_{N-2}&0&0\\
  1 &  p_1(y)&\ldots& p_{N-2}(y)& p_{N-1}(y) & p_{N}(y) \\
  1&p_1(x)&\ldots&p_{N-2}(x)& p_{N-1}(x)&p_{N}(x)
 \end{array}\right].
\]
The Christoffel-Darboux is established by comparing the last
expression with Eq.\ \eqref{CD1}:
\begin{equation}
\sum_{n=1}^{N}\frac{1}{h_{n-1}}p_{n-1}(y)p_{n-1}(x)=\frac{1}{h_{N-1}}\frac{p_N(x)p_{N-1}(y)-p_{N-1}(x)p_N(y)}{x-y}.\end{equation}

\section{Multiple polynomials of type I and II}

Let us first review some properties of multiple polynomials (e.g.,
see \cite{Aptekarev, DK, VAC}).    These mathematical objects are
associated to $D$ distinct weight functions $w_i(x)$. (We limit
ourself to the so-called AT systems in which the support $I$ is the
same for all weights $w_i$.) They are indexed by
$\vec{n}=(n_1,\ldots,n_D)$,  a composition (or a multi-index) of
length $D$ and of weight $N$; that is, an ordered sequence of $D$
non-negative integers $n_i$ such that
$|\vec{n}|:=\sum_{i=1}^Nn_i=N$.

To each composition $\vec{n}$, we associate $D$ multiple polynomials
of type I, here denoted by $A^{(i)}_{\vec{n}}$, where $i$ varies
from $1$ to $D$. They are generated by the multiple
function
\begin{equation}Q_{\vec{n}}(x)=\vec{w}(x)\cdot\vec{A}_{\vec{n}}(x):=\sum_{i=1}^Dw^{(i)}(x)A^{(i)}_{\vec{n}}(x).
\end{equation}
This function  satisfies a simple orthogonality condition
\begin{equation}\label{OrthoI}\int_Idx\,x^{j}Q_{\vec{n}}(x)=\begin{cases}
0,&j=0,\ldots,|\vec{n}|-2,\\
1,&j=|\vec{n}|-1.\end{cases}
\end{equation}
 Note that the degree
of $A^{(i)}_{\vec{n}}$ is assumed to be $n_i-1$ (technically
speaking, we only work with perfect systems).

The multiple polynomial of type II characterized by the composition
$\vec{n}$ is denoted by $P_{\vec{n}}$.  It is a monic polynomial of
weight $|\vec{n}|$ that complies with $D$ orthogonality relations,
\begin{equation}\label{OrthoII}
\int_Idx\,w^{(i)}(x)x^jP_{\vec{n}}(x)=0,\qquad
j=0,\ldots,n_i-1,\qquad i=1,\ldots,D.\end{equation}

The multiple functions $Q_{\vec{n}}$ and $P_{\vec{n}}$ provide a
biorthogonal system.  Indeed, first fix $\vec{n}=(n_1,\ldots,n_D)$
with $|\vec{n}|=N$. Second, choose a sequence of compositions such
that \[|\vec{n}^{(i)}|=i\qquad\mbox{ and} \qquad\vec{n}^{(i)}_j\leq
\vec{n}^{(i+1)}_j\] for all $ i=0,\ldots,N-1$ and $j=1,\ldots,D$.
For instance, one could take\[\begin{array}{ll}
\vec{n}^{(0)}&=({0,0,0,\ldots}),\\
\vec{n}^{(1)}&=({1,0,0,\ldots}),\\
&\vdots \\
\vec{n}^{(n_1)}&=({n_1,0,0,\ldots}),\\
\vec{n}^{(n_1+1)}&=({n_1,1,0,\ldots}),\\
&\vdots\\
\vec{n}^{(N)}&=({n_1,n_2,\ldots,n_D}). \end{array}
\]
Third, define
\begin{equation}
P_i:=P_{\vec{n}^{(i)}},\qquad
Q_i:=Q_{\vec{n}^{(i+1)}},\qquad i=0,\ldots,N-1.
\end{equation}
 Then, we see from relations \eqref{OrthoI} and \eqref{OrthoII} that these functions are
biorthogonal:
\begin{equation}
\int_Idx\,P_{i}(x)Q_{j}(x)=\delta_{i,j} \qquad \mbox{for all}\qquad
i,j=0,\ldots,N-1.
\end{equation}

\begin{proposition}
Suppose that we have a matrix ensemble with eigenvalue p.d.f. of the
form \eqref{biorthopdf} with $\eta_{i}(x)=x^{i-1}$, or 
equivalently
a monic polynomial of degree $i-1$, and
\begin{multline}\label{multiweights}
\big[\xi_1,\xi_2,\ldots,\xi_{N}\big](x)=\big[w^{(1)}(x),xw^{(1)}(x),\ldots,x^{n_1-1}w^{(1)}(x),w^{(2)}(x),xw^{(2)}(x),\\
\ldots,x^{n_2-1}w^{(2)}(x),\ldots
w^{(D)}(x),xw^{(D)}(x),\ldots,x^{n_D-1}w^{(D)}(x)\big].
\end{multline}
%\begin{align*}\big(\xi_1,\xi_2,\ldots,\xi_{n_1}\big)(x)&=\big(w_1(x),xw_1(x),\ldots,x^{n_1-1}w_1(x)\big)\\
%\big(\xi_{n_1+1},
%\xi_{n_1+2},\ldots,\xi_{n_1+n_2}\big)(x)&=\big(w_2(x),xw_2(x),\ldots,x^{n_2-1}w_2(x)\big)\\
%&\;\;\vdots\\
%\big(\xi_{N-n_D},
%\xi_{N-n_D+1},\ldots,\xi_{N}\big)(x)&=\big(w_D(x),xw_D(x),\ldots,x^{n_D-1}w_D(x)\big).
%\end{align*}
Suppose moreover that
$\mathbf{g}:=[g_{i,j}]_{i,j=1}^N$ is
non-singular, where $g_{i,j}=\int_Idx\,\eta_i(x)\xi_j(x)$. Let
\begin{equation}\label{Qaverage}
Q_{\vec{n}}(x)=\Res{z=x}\left\langle{\det(z\II-\mathbf{X})}^{-1}\right\rangle,
\end{equation}
for $z\in\mathbb{C}\setminus\mathbb{R}$, and
\begin{equation}\label{Paverage}
P_{\vec{n}}(x)=\langle\det(x\II-\mathbf{X})\rangle.
\end{equation}
Then
\begin{align}\label{detQ}
Q_{\vec{n}}(x)&=\frac{N!}{Z_N}\left|\begin{array}{cccc}
                                     g_{1,1} & g_{1,2}&\ldots & g_{1,N} \\
                                     \vdots& \vdots&\ddots & \vdots \\
                                     g_{N-1,1}& g_{N-1,2}&\ldots & g_{N-1,N} \\
                                     \xi_1(x) & \xi_2(x)&\ldots & \xi_N(x)
                                   \end{array}\right|,\\\label{detP}
P_{\vec{n}}(x)&=\frac{N!}{Z_N}\left|\begin{array}{cccc}
                                     g_{1,1}\, &\ldots & g_{1,N}&\eta_1(x) \\
                                      g_{2,1}\, &\ldots & g_{2,N}&\eta_2(x) \\
                                     \vdots& \vdots&\vdots & \vdots \\
                                     g_{N+1,1}\,&\ldots& g_{N+1,N}&\eta_{N+1}(x) \\
                                                                       \end{array}\right|,
 \end{align} where
 $Z_N=N!\det[g_{i,j}]_{i,j=1}^N$. Furthermore, $Q_{\vec{n}}$ and
$P_{\vec{n}}$ are the only functions satisfying Eqs.\ \eqref{OrthoI}
and \eqref{OrthoII}.
\end{proposition}
\begin{proof}
In \cite{BKmulti}, Bleher and Kuijlaars have shown that
\eqref{OrthoII} holds true if the type II polynomials are given by
\eqref{Paverage}, or equivalently by \eqref{detP}, with
$w^{(j)}(x)=w(x)x^{d_j-1}e^{a_i x}$ where
$d_j=j-\sum_{k=1}^{i-1}n_k$, for $i$ such that
$\sum_{k=1}^{i-1}n_k<j\leq \sum_{k=1}^{i}n_k$. The generalization of
their argument to our case is immediate. So, let us concentrate on
type I functions, defined in Eq.\ \eqref{Qaverage}.

Firstly, by following the method exposed in the proof of Proposition
1, we find
\begin{align*}
Q_{\vec{n}}(x)&=\frac{1}{Z_N}\,\Res{z=x}\,\int_Idx_1\cdots\int_Idx_N\det[\xi_i(x_j)]_{i,j=1}^N\det[x_j^{i-1}]_{i,j=1}^N\prod_{i=1}^N\frac{1}{z-x_i}\\
&=\frac{N!}{Z_N}\,\Res{z=x}\,\int_Idx_1\,\xi_1(x_1)\cdots\int_Idx_N\,\xi_N(x_N)\det\left[\begin{array}{c}
\big[\eta_i(x_j)\big]_{\substack{i=1\ldots{N-1}\\j=1,\ldots,N}}\\
\big[\frac{1}{z-x_j}\big]_{j=1,\ldots,N} \end{array}\right]\\
&=\frac{N!}{Z_N}\,\Res{z=x}\,\int_Idx_1\cdots\int_Idx_N\det\left[\begin{array}{c}
\big[x_j^{i-1}\xi_j(x)\big]_{\substack{i=1\ldots{N-1}\\j=1,\ldots,N}}\\
\big[\frac{x_j^{i-1}}{z-x_j}\big]_{j=1,\ldots,N} \end{array}\right].
\end{align*} The last line obviously leads to Eq.\ \eqref{detQ}.

Secondly, we choose $\eta_i(x)=x^{i-1}$ and set
\begin{equation}
h^{(i)}_j:=\int_Idx\,w_i(x)\eta_j(x),
\end{equation} so that the determinantal expression of
$Q_{\vec{n}}$ becomes
\begin{equation}
Q_{\vec{n}}(x)=\frac{N!}{Z_N}
\left|\begin{array}{cccc}
 \big[h^{(1)}_{i+j-1}]_{\substack{i=1,\ldots,N-1\\j=1\ldots,n_1}} & \big[h^{(2)}_{i+j-1}]_{\substack{i=1,\ldots,N-1\\j=1\ldots,n_2}}&\ldots & \big[h^{(D)}_{i+j-1}]_{\substack{i=1,\ldots,N-1\\j=1\ldots,n_D}} \\
  \big[\eta_j(x)w_1(x)\big]_{j=1,\ldots,n_1} & \big[\eta_j(x)w_2(x)\big]_{j=1,\ldots,n_2} &\ldots & \big[\eta_j(x)w_D(x)\big]_{j=1,\ldots,n_D}
  \end{array}\right|.
\end{equation}
Hence,
\begin{equation*}
\int_Idx\,x^{k} Q_{\vec{n}}(x)=\frac{N!}{Z_N}
\left|\begin{array}{cccc}
 \big[h^{(1)}_{i+j-1}]_{\substack{i=1,\ldots,N-1\\j=1\ldots,n_1}} & \big[h^{(2)}_{i+j-1}]_{\substack{i=1,\ldots,N-1\\j=1\ldots,n_2}}&\ldots & \big[h^{(D)}_{i+j-1}]_{\substack{i=1,\ldots,N-1\\j=1\ldots,n_D}} \\
  \big[h^{(1)}_{j+k}\big]_{j=1,\ldots,n_1} & \big[h^{(2)}_{j+k}\big]_{j=1,\ldots,n_2} &\ldots & \big[h^{(D)}_{j+k}\big]_{j=1,\ldots,n_D}
  \end{array}\right|.
\end{equation*}
We see that the r.h.s.\ is null when the last row equals one of
other the rows; i.e., when $k=0,\ldots,N-2$.  For $k=N-1$, the
determinant simply becomes $\det [g_{i,j}]_{i,j=1}^N=Z_N/N!$. This
completes the proof of the orthogonality condition \eqref{OrthoI}.

We finally show the uniqueness of the type II multiple function.
Expression \eqref{detQ} tells us that
$Q_{\vec{n}}(x)=\sum_{i=1}w^{(i)}(x)A^{(i)}_{\vec{n}}(x)$, where
$A^{(i)}_{\vec{n}}=c^{(i)}_1x^{n_i-1}+c^{(i)}_2x^{n_i-2}+\ldots+c^{(i)}_{n_i}$.
This means that, in order to determine $Q_{\vec{n}}$ uniquely, we
have to fix the $N=|\vec{n}|$ coefficients $c^{(i)}_j$
($j=1,\dots,n_i$, $i=1,\ldots,D$).  But Eq.\ \eqref{OrthoI}
furnishes exactly $N$ linear equations.  The matrix for that linear
system is
\begin{equation}
\mathbf{g}=\left[\begin{array}{ccc}
                  g_{1,1} & \ldots & g_{1,N} \\
                  \vdots & \ddots & \vdots \\
                  g_{N,1} & \ldots & g_{N,N}
                \end{array}\right]
=\left[\begin{array}{cccc}
 \big[h^{(1)}_{i+j-1}]_{\substack{i=1,\ldots,N\\j=1\ldots,n_1}} & \big[h^{(2)}_{i+j-1}]_{\substack{i=1,\ldots,N\\j=1\ldots,n_2}}&\ldots & \big[h^{(D)}_{i+j-1}]_{\substack{i=1,\ldots,N\\j=1\ldots,n_D}} \\
    \end{array}\right].
\end{equation}
But by hypothesis $\det \mathbf{g}\neq 0$. Consequently, the
solution for the coefficients, and therefore $Q_{\vec{n}}$, is
unique.
 \end{proof}

\section{Perturbation of chiral unitary ensembles}

As we mentioned in the introduction, non-trivial biorthogonal matrix
ensembles exist. For instance, ensembles of $N\times N$ Hermitian
matrix with a p.d.f.\ proportional to $\exp(-\tr V(\XX)+\tr
\mathbf{A}\XX)$, where $\mathbf{A}$ is a fixed $N\times N$ Hermitan
matrix, naturally lead to biorthogonal systems.  Specifically, let
$\xi_i(x)=\exp(-V(x)+a_ix)$ and let $\mathbf{a}:=\{a_1,\ldots,a_N\}$
denote the eigenvalues of $\mathbf{A}$.  Suppose that some of the
$a_i$'s coincide, i.e.,
\begin{equation}\label{limeigen}
a_{n_1}\rightarrow a_{n_1-1}\rightarrow \ldots\rightarrow
a_{1}=b_1,\qquad a_{n_1+n_2}\rightarrow
a_{n_2+n_1-1}\rightarrow\ldots\rightarrow a_{n_1+1}=b_2,
\end{equation}
and so on. Symbolically, this is written as
\begin{equation}
\mathbf{a}=\mathbf{b}^{\vec{n}}.
\end{equation}
Then from the definition of the function $\xi_i$ one sees 
\begin{equation}
\xi_j(x)=\xi_i(x)\sum_{n\geq 0}\frac{(a_j-a_i)^n}{n!}x^n,
\end{equation}
uniformly for $|a_j-a_i|<\infty$, so that \cite{BKmulti}
\begin{equation}
\lim \frac{\det[\xi_i(x_j)]_{i,j=1}^N}{\prod_{1\leq i<j\leq N}(a_j-a_i)}=\frac{\det[\bar{\xi}_i(x_j)]_{i,j=1}^N}{\prod_{i=1}^D\prod_{j=1}^{n_i-1}j!\prod_{1\leq k<\ell\leq D}(b_\ell-b_k)^{n_k n_\ell}}
\end{equation}
where $[\bar{\xi}_1,\ldots,\bar{\xi}_N]$ is given by Eq.\
\eqref{multiweights}, with $w_i(x)=\exp(-V(x)+b_i x)$. Therefore,
these biorthogonal ensembles can be studied with the help of
multiple polynomials (see Proposition 2).  In this section, we
provide other examples of biorthogonal matrix ensembles. This time,
weight functions are of the form
$w^{(i)}(x)=x^{\alpha}e^{-V(x)}I_{\alpha}(2\sqrt{xb_i})$, where
$I_\alpha$ is the modified Bessel function (see below).

Suppose $\alpha:=M-N\geq 0$. Let $\XX=[X_{i,j}]$, be a random
$M\times N$ (non-Hermitian) complex matrix drawn with probability
\begin{equation}\label{pdfchiral}
P(\XX)(d\XX)= e^{-\tr
V(\XX^\dagger\XX)}e^{\RRe(\tr\XX\mathbf{A}^{\dagger})}(d\XX).
\end{equation}
$\mathbf{A}$ is a fixed $M\times N$ complex matrix and $(d\XX)$
denotes the normalized volume element of $\mathbb{C}^{M\times N}$,
\[(d\XX):=\frac{1}{C}\prod_{i=1}^M\prod_{j=1}^Nd\RRe(X_{i,j})d\IIm(X_{i,j}).\]
The "potential" $V$ has to be chosen in a way that guarantees the
positivity of $\tr V(\XX)$.  When $\mathbf{A}=\mathbf{0}_{M\times
N}$ and $V(x)=x$, the p.d.f.\ \eqref{pdfchiral} defines the chiral
Unitary Ensemble (chUE), which is simply related to the Laguerre
Unitary Ensemble (LUE) (for more details, see \cite[Chapter
2]{ForresterBook}).

We now want to get the eigenvalue p.d.f.\ associated to
\eqref{pdfchiral}.  This can be done through a singular value
decomposition of $\XX$: \begin{equation}\label{tansform}
\XX=\tilde{\mathbf{U}}\XX_\mathrm{D}\tilde{\mathbf{V}}^\dagger,
\end{equation}
where
\begin{equation} \XX_\mathrm{D}=\left[
\begin{array}{c}
\mathrm{diag}[s_1,\ldots,s_N]\\
\mathbf{0}_{\alpha\times N} \end{array}\right],
\quad\tilde{\mathbf{U}}\in U(M),\quad\tilde{\mathbf{V}}\in
U(N).\end{equation}Note that the singular values $s_1,\ldots, s_N$
are real and non-negative; they are the positive square roots of the
eigenvalues of the $N\times N$ matrix $\XX^\dagger\XX$.  A similar
decomposition is possible for the non-random matrix; that is,
$\mathbf{A}=\bar{\mathbf{U}}\mathbf{A}_\mathrm{D}\bar{\mathbf{V}}^\dagger$
with
$\mathbf{A}_\mathrm{D}^\dagger=\left[\mathrm{diag}[t_1,\ldots,t_N]\,\mathbf{0}_{N\times
\alpha}\right]$ and $t_1,\ldots,t_N\geq 0$. For the moment, we
assume $t_i\neq t_j$ for $i\neq j$.  For notational convenience we set
\begin{equation} s_i^2=x_i,\qquad t_i^2=4a_i,\qquad
i=1,\ldots,N.\end{equation} By considering the transformation
$\eqref{tansform}$, and its Jacobian, we get an integral form for the
eigenvalue p.d.f.
\begin{multline}
p_N(x_1,\ldots,x_N)\propto \prod_{i=1}^Nx_i^{\alpha}
e^{-V(x_i)}\prod_{1\leq i<j\leq N}(x_j-x_i)^{2}\\
\times\int_{U(N)}(\mathbf{V}^{\dagger}d\mathbf{V})\int_{U(M)}(\mathbf{U}^{\dagger}d\mathbf{U})
\exp \left\{\RRe (\tr\,
\XX_{\mathrm{D}}\mathbf{V}^{\dagger}\mathbf{A}_{\mathrm{D}}^{\dagger}\mathbf{U})\right\},
\end{multline}
where $(\mathbf{U}^{\dagger}d\mathbf{U})$ is, up to a multiplicative
factor, the Haar measure on the unitary group $U(M)$ (and similarly for $\mathbf{V}$).
The integration can be realized by making use of a Itzykson-Zuber
type formula \cite{JSV,ZZ}:
\begin{multline}
\int_{U(N)}(\mathbf{V}^{\dagger}d\mathbf{V})\int_{U(M)}(\mathbf{U}^{\dagger}d\mathbf{U})
\exp \left\{\RRe (\tr\,
\XX_{\mathrm{D}}\mathbf{V}^{\dagger}\mathbf{A}_{\mathrm{D}}^{\dagger}\mathbf{U})\right\}=\\
C_{M,N}\prod_{i=1}^N\frac{1}{(a_ix_i)^{\alpha/2}}\frac{\det\big[I_\alpha(
2\sqrt{a_ix_j})\big]_{i,j=1}^N
}{\Delta_N(a_1,\ldots,a_N)\Delta_N(x_1,\ldots,x_N)},
\end{multline}
where $C_{M,N}$ is a constant independent of the $x_i$'s and
$a_i$'s.  Recall that the modified Bessel function of the first kind
is specified by 
\begin{equation}\label{BesselI}
I_\alpha(z)=I_{-\alpha}(z)=\left(\frac{z}{2}\right)^\alpha\sum_{k\geq
0}\frac{(z^2/4)^k}{\Gamma(k+1)\Gamma(\alpha+k+1)}=
\int_{\mathcal{C}_{\{0\}}}\frac{dw}{2\pi\im}\frac{e^{z/2(w+w^{-1})}}{w^{\pm\alpha+1}},
\end{equation}
where it is assumed that $\alpha\in\mathbb{Z}$, and where
$\mathcal{C}_{\{0\}}$ stands for a positive contour that encircles
the origin.  It can be expressed as a hypergeometric function as
well,
\begin{equation}\label{I0F1}
I_{\alpha}(2z^{1/2})=\frac{z^{\alpha/2}}{\Gamma(\alpha+1)}
\,_0F_1(\alpha+1,z),\qquad \alpha\in\mathbb{C},\qquad |\arg(z)|<\pi.
\end{equation}
Combining the few last equations, we obtain the eigenvalue (or
singular value) p.d.f.\
\begin{equation}\label{epdf}
p_N(x_1,\ldots,x_N)=\frac{1}{Z'_N}\prod_{i=1}^Nx_i^{\alpha}
e^{-V(x_i)}\prod_{1\leq i<j\leq
N}\left(\frac{x_j-x_i}{a_j-a_i}\right)\det\left[
\frac{\,_0F_1(\alpha+1,a_ix_j)}{\Gamma(\alpha+1)}\right]_{i,j=1}^N.
\end{equation}
L'Hospital's rule provides the appropriate density when some of the
$a_i$'s coincide.  Clearly, Eq.\ \eqref{epdf} is of the biorthogonal
form, with
\begin{equation}
\eta_{i}(x)=x^{i-1}+\mbox{lower terms},\qquad
\xi_i(x)=\frac{x^{\alpha}e^{-V(x)}}{\Gamma(\alpha+1)}\,_0F_1(\alpha+1,a_ix).
\end{equation}
As a consequence, the correlation functions satisfy
$\rho_{n,N}(x_1,\ldots,x_n)=\det\left[K_N(x_i,x_j)\right]_{i,j=1}^n$
and the kernel is given by $(x-y)K_N(x,y)=\Res{z=y}\,
\langle{\det(x\II-\mathbf{X})}{\det(z\II-\mathbf{X})}^{-1} \rangle$.

When we perturb  ensembles of Hermitian matrices by a source term,
 the multiple functions $Q_{\vec{n}}$ and $P_{\vec{n}}$ can be obtained through Proposition 2.
 In that case, the composition $\vec{n}=(n_1,\ldots,n_D)$ gives the multiplicity of the eigenvalues
 $(b_1,\ldots,b_D)$ (see limit \eqref{limeigen}).  One might be tempted to conclude that this relation
 remains the same in perturbed ensembles of rectangular complex matrices.  It is true that Proposition 2
 still holds. However, the link between the multi-index $\vec{n}$ and the eigenvalues $ (b_1,\ldots,b_D)$,
 or equivalently between the function $\xi_i$ and the weight functions $w_i$,  is more involved.
 The following lemma and proposition aim to clarify the situation.

\begin{lemma} Let $\xi_i(x)=w_\alpha (x,a_i)$, where
\begin{equation} \label{walpha}
w_\alpha(x,a_i):=\frac{x^{\alpha}e^{-V(x)}}{\Gamma(\alpha+1)}\,_0F_1(\alpha+1,a_ix).
\end{equation}Consider the limit \eqref{limeigen}.   Then,
\begin{equation}\label{eqlimdet}
\lim \frac{\det[\xi_i(x_j)]_{i,j=1}^N}{\prod_{1\leq i<j\leq
N}(a_j-a_i)} =\frac{\det\left[ W^{(1)}_\alpha\quad
W^{(2)}_\alpha\quad
 \ldots\quad
W^{(D)}_\alpha\right]}{\prod_{i=1}^D\prod_{j=1}^{n_i-1}j!\prod_{1\leq
k<\ell\leq D}(b_\ell-b_k)^{n_k n_\ell}},
\end{equation}
where \[W^{(k)}_\alpha=\Big[w_\alpha(x_i,b_k)\quad w_{\alpha+1}(
x_i,b_k)\quad\ldots\quad
w_{\alpha+n_k-1}(x_i,b_k)\Big]_{i=1,\ldots,N}.\]
\end{lemma}
\begin{proof}
First, we suppose that, as $a_n\rightarrow a_{n-1} \rightarrow
\ldots\rightarrow a_1=b_1$, the following equation holds true
\begin{multline}\label{formulalim}
G_n:=\lim {\prod_{1\leq i<j\leq
N}(a_j-a_i)^{-1}}{\det[\xi_i(x_j)]_{i,j=1}^N}=\\
\prod_{k=1}^{n-1}(k!)^{-1}{\prod_{i=n+1}^N(a_i-b_1)^{-n}\prod_{n+1\leq
i<j\leq N}(a_j-a_i)^{-1}} \times\left| \begin{array}{l}
\big[w_{\alpha+i-1}(x_j,b_1)\big]_{i=1,\ldots,n}\\
\big[w_{\alpha}(x_j,a_i)\big]_{i=n+1,\ldots,N}
\end{array}
\right|_{i=1,\ldots,N}.
\end{multline}
Second, we consider the series expansion of $I_\alpha$, given by
Eq.\ \eqref{BesselI}, from which we deduce
\begin{equation} \label {formulaBessel}
w_\alpha(x,a_j)=\sum_{\ell\geq 0}\frac{(a_j-a_i)^\ell}{\ell!}w_{\alpha+\ell}(x,a_i)
\end{equation}
uniformly for $|a_j-a_i|<\infty$.  We thus have
 \begin{multline*}
\lim_{a_2\rightarrow a_1} {\prod_{1\leq i<j\leq
N}(a_j-a_i)^{-1}}{\det[\xi_i(x_j)]_{i,j=1}^N}\\
={\prod_{i=3}^N(a_i-b_1)^{-2}\prod_{3\leq
i<j\leq N}(a_j-a_i)^{-1}}\lim_{a_2\rightarrow a_1} (a_2-a_1)^{-1}\\
\times \left|
\begin{array}{c}
\big[w_{\alpha}(x_j,a_1)\big] \\
 \big[w_{\alpha}(x_i,a_1) +(a_2-a_1)w_{\alpha+1}(x_i,a_1)+O\left((a_2-a_1)^2\right)\big]\\
 \big[w_\alpha(x_j,a_{i})\big]_{i=3,\ldots,N}
 \end{array}\right|_{j=1,\ldots,N}.
\end{multline*}
But we can subtract the first row from second without affecting the
determinant, so that
\begin{multline*}
G_2=\lim_{a_2\rightarrow a_1=b_1}
\frac{\det[\xi_i(x_j)]_{i,j=1}^N}{\prod_{1\leq i<j\leq
N}(a_j-a_i)}\\
= {\prod_{i=2}^N(a_2-a_i)^{-2}\prod_{3\leq i<j\leq
N}(a_j-a_i)^{-1}}
\times\left|\begin{array}{l}\big[w_{\alpha}(x_j,b_1)\big]\\
\big[w_{\alpha+1}(x_j,b_1)\big]\\
\big[w_\alpha(x_j,a_i)\big]_{i=3\ldots N}
\end{array}\right|_{j=1,\ldots,N}.
\end{multline*}
This shows Eq.\ \eqref{formulalim} for $n=2$.  The general $n$ case
is established by induction: we return to \eqref{formulalim}; we
apply \eqref{formulaBessel} once again, i.e.,
\begin{multline*}
\lim_{a_{n+1}\rightarrow{b_1}}
G_n=\prod_{k=1}^{n-1}(k!)^{-1}{\prod_{i=n+2}^N(a_n-a_i)^{-n-1}\prod_{n+2\leq
i<j\leq N}(a_j-a_i)^{-1}}\\
\lim_{a_{n+1}\rightarrow{b_1}}(a_{n+1}-b_1)^{-n}
\times\left|\begin{array}{c}\big[w_{\alpha+i-1}(x_j,b_1)\big]_{i=1,\ldots,n}\\
\big[\sum_{k\geq 0}(a_{n+1}-b_1)^{k}(k!)^{-1}w_{\alpha+k}(x_j,b_1)\big]\\
\big[w_{\alpha}(x_j,a_{i})\big]_{i=n+2,\ldots, N}
\end{array}\right|_{j=1\ldots N};
\end{multline*}
 we manipulate the rows as,
\[ \row (n+1) \rightarrow \row (n+1) - \row (1)-(a_{n+1}-b_1)\row (2)-\ldots
-\frac{(a_{n+1}-b_1)^{n-1}}{(n-1)!}\row (n);\] and we finally get
\[\prod_{k=1}^{n}(k!)^{-1}{\prod_{i=n+2}^N(a_n-a_i)^{-n-1}\prod_{n+2\leq
i<j\leq N}(a_j-a_i)^{-1}}
\times\left|\begin{array}{c}\big[w_{\alpha+i-1}(x_j,b_1)\big]_{i=1,\ldots,n+1}\\
\big[w_{\alpha}(x_j,a_{i})\big]_{i=n+2,\ldots, N}
\end{array}\right|_{j=1\ldots N},\]which is $G_{n+1}$, as expected.
The general formula \eqref{eqlimdet} is obtained by taking $D$
limits similar to \eqref{formulalim}.
\end{proof}

\begin{proposition} Consider the functions $\xi_i$ and $w_\alpha$ as defined in the previous lemma. Let $w_\alpha(x)=w_\alpha(x,0)$.
 Suppose
\begin{equation}\label{limeigen2}
a_r\rightarrow a_{r-1}\rightarrow \ldots\rightarrow a_{1}=b>0,\qquad
a_N\rightarrow a_{N-1}\rightarrow \ldots\rightarrow a_{r+1}=0,
\end{equation}
Then,
\begin{multline}
\lim \frac{\det[\xi_i(x_j)]_{i,j=1}^N}{\prod_{1\leq i<j\leq
N}(a_j-a_i)} =\\
\frac{(-b)^{-r(N-r)}\prod_{s=1}^{\lfloor\frac{r-1}{2}\rfloor}(-b)^{-s}\prod_{t=1}^{\lfloor\frac{r+1}{2}\rfloor}b^{-t}}{\prod_{k=1}^{r-1}
k!\prod_{\ell=1}^{N-r-1}\ell!} \left|\begin{array}{l}
\Big[x_j^{i-1}w_\alpha(x_j)\Big]_{i=1,\ldots,N-r}\\
\Big[x_j^{i-1}w_\alpha(x_j,b)\Big]_{{i=1,\ldots \lfloor
\frac{r+1}{2}\rfloor}}\\
\Big[x_j^{i-1}w_{\alpha+1}(x_j,b)\Big]_{{i=1,\ldots \lfloor
\frac{r-1}{2}\rfloor}}
\end{array}
\right|_{j=1,\ldots,N}.
\end{multline}
\end{proposition}
\begin{proof}
We have from the previous lemma and from $w_{\alpha+i}(x,0)=x^{i}w_\alpha(x)$that
 \[ \lim \frac{\det[\xi_i(x_j)]_{i,j=1}^N}{\prod_{1\leq i<j\leq
N}(a_j-a_i)}=\frac{1}{(-b)^{r(N-r)}\prod_{k=1}^{r-1}
k!\prod_{\ell=1}^{N-r-1}\ell!} \left|\begin{array}{l}
\Big[x_j^{i-1}w_{\alpha}(x_j)\Big]_{\substack{i=1,\ldots,N-r\\j=1,\ldots,N}}\\
\Big[w_{\alpha+i-1}(x_j,b)\Big]_{\substack{i=1,\ldots
r\\j=1,\ldots,N}}
\end{array}
\right|.\] Now, for $\alpha>-1$, it is known that
\[x\,_0F_1(\alpha+3,x)=(\alpha+1)(\alpha+2)\Big(\,_0F_1(\alpha+1,x)-\,_0F_1(\alpha+2,x)\Big).\]
This implies for $b\neq 0$ and $k\in\mathbb{N}$,
\[b\,w_{\alpha+k}(x,b)=x\,w_{\alpha+k-2}-(\alpha+k-1)w_{\alpha+k-1}(x,b).\]The latter identity allows us to write
\[w_{\alpha+k}(x)=
\left(\frac{x}{b}\right)^{\lfloor\frac{k+1}{2}\rfloor}w_{\alpha+2\left(\frac{k}{2}-\lfloor
\frac{k}{2}\rfloor\right)}(x) +c_\alpha(x;b,k)\] where $c_\alpha$ is
a linear combination of $w_\alpha$ and $w_{\alpha+1}$ with
coefficient depending on $b$ and $k$.  The desired result is
obtained by using the latter equation and by exploiting the
antisymmetry of the determinant under the permutation of the rows as
well as the invariance of the determinant under the transformation
$\row(i)\rightarrow\row(i)+\sum_{j\neq i}c_{i,j}\row(j)$.
\end{proof}

The last proposition implies that each limit of the form
$a_{n_1}\rightarrow a_{n_1-1}\rightarrow \ldots \rightarrow a_1=b>0$
give rise to two functionally independent weight functions, i.e.,
$w_\alpha(x,b)$ and $w_{\alpha+1}(x,b)$. When we have $D$ similar
limits, with $b_1>b_2>\ldots>b_D$ say, we get $2D$ weight functions
if $b_D>0$, and $2D-1$ weight functions if $b_D=0$.

\section{Chiral Gaussian Unitary Ensemble with a source term}

In the next paragraphs, we focus on the perturbation of the chGUE. So
we choose
\begin{equation}\label{defch2} V(x)=x,\qquad
\eta_k(x)=(-1)^{k-1}(k-1)!L^{\alpha}_{k-1}(x),\qquad
\xi_i(x)=\frac{x^{\alpha}e^{-x}\,_0F_1(\alpha+1,a_ix)}{\Gamma(\alpha+1)}\end{equation}
where $L^{\alpha}_{k}$ denotes the (associated) Laguerre polynomial
of degree $k$.
\begin{proposition}  Let $\mathbf{a}=\{a_1,\ldots,a_N\}\in (0,\infty)^N$.  Then 
the kernel of the perturbed chGUE, as defined by Eqs.\ \eqref{pdfchiral} and \eqref{defch2},  is given by
\begin{equation}\label{Kfirst}
K_N(x,y)=\frac{y^\alpha e^{-y+x}}{\Gamma(\alpha+1)^2}\!\int_0^\infty
du\,u^\alpha e^{-u}\,_0
F_1(\alpha+1,-xu)\!\!\int_{\mathcal{C}^{\{u\}}_{-\mathbf{a}}}\!\!\frac{dv}{2\pi\im}
\frac{e^v\,_0F_1(\alpha+1,-yv)}{u-v}\prod_{i=1}^N\frac{u+a_i}{v+a_i},\end{equation}
where $\mathcal{C}^{\{u\}}_{-\mathbf{a}}$ denotes a counterclockwise
contour encircling the points $-a_1,\ldots,-a_N$ but not the point
u.  Equivalently,
\begin{multline}\label{Ksecond}
\frac{w_\alpha(x)}{w_\alpha(y)}\,K_N(x,y)=\\
 \int_0^\infty du \int_{\mathcal{C}^{\{u\}}_{\mathbf{b}}}
\frac{dv}{2\pi\im}\int_{\mathcal{C}_{\{0\}}}\frac{dw}{2\pi\im}
\int_{\mathcal{C}_{\{0\}}}\frac{dz}{2\pi\im}
\frac{e^{v-u}}{u-v}\frac{e^{-u/w+v/z}}{zw}{e^{xw-yz}}\left(\frac{u}{v}\right)^\alpha
\left(\frac{z}{w}\right)^\alpha\prod_{i=1}^N\frac{u+a_i}{v+a_i},
\end{multline}
where $\mathcal{C}_{\{0\}}$ is a positive contour around the origin.
When some of the parameters $a_i$'s are null, the previous equations
remain valid if $\int_0^\infty du$ is understood as
$\lim_{\epsilon\rightarrow  0^+}\int_\epsilon^\infty du$.
\end{proposition}
\begin{proof}
In the first steps, we assume that all the elements of $\mathbf{b}$
are distinct.   As mentioned in the introduction (see also
Proposition 1), the kernel can be written as
$K_N(x,y)=\sum_{i,j=1}^N \eta_i(x)c_{i,j}\xi_{j}(y)$ with
$[c_{j,i}]=[g_{i,j}]^{-1}$ ($\mathbf{c}^\TT=\mathbf{g}^{-1}$).
Explicitly,
\begin{align}\nonumber
K_N(x,y)&=\frac{y^\alpha e^{-y}}{\Gamma(\alpha+1)}\sum_{i,j=1}^N (-1)^{i-1}(i-1)!L^{\alpha}_{i-1}c_{i,j}\,_0F_1(\alpha+1,a_jy)\\
&=\frac{y^\alpha e^{-y+x}}{\Gamma(\alpha+1)^2}\sum_{i,j=1}^N
(-1)^{i-1}c_{i,j}\int_0^\infty
du\,u^{\alpha+i-1}e^{-u}\,_0F_1(\alpha+1,-xu)\,_0F_1(\alpha+1,a_jy)\label{Ksumint}
\end{align}
where we have made use of the formula (cf.\ \cite[Eq.\ (5.4.1)]{Szego} and Eq.\ \eqref{I0F1} above)
\begin{equation}\label{L0F1}
L^{\alpha}_{n}(x)=\frac{e^x}{n!\Gamma(\alpha+1)}\int_0^\infty
du\,u^{\alpha+n}e^{-u}\,_0F_1(\alpha+1,-xu).
\end{equation}
We now aim to eliminate the  coefficient $c_{i,j}$.  In our case,
\begin{equation}
g_{i,j}=\frac{(-1)^{i-1}(i-1)!}{\Gamma(\alpha+1)}\int_{0}^\infty
dx\,x^{\alpha}e^{-x}L^{\alpha}_{i-1}(x)\,_0F_1(\alpha+1,a_j x).
\end{equation}
This can be evaluated exactly:
\begin{equation}
g_{i,j}=a_j^{i-1} e^{a_j}.
\end{equation}
As a consequence,  $c_{i,j}$ must comply with
$e^{a_k}\sum_{i=1}^N(a_k)^{i-1}c_{i,j}=\delta_{j,k}$.  Thus
\begin{equation}
\sum_{i=1}^N (-u)^{i-1}
c_{i,j}=(-1)^{N-1}e^{-a_j}\prod_{\substack{\ell =1\\\ell\neq
j}}^N\frac{u+a_\ell}{a_j-a_\ell}.
\end{equation}
By substituting the last equation into \eqref{Ksumint}, we find
\begin{equation}
K_N(x,y)=\frac{y^\alpha e^{-y+x}}{\Gamma(\alpha+1)^2}\int_0^\infty
du\,u^{\alpha}e^{-u}\,_0F_1(\alpha+1,-xu)\sum_{j=1}^N
\,_0F_1(\alpha+1,a_jy)e^{-a_j}\prod_{\substack{\ell =1\\\ell\neq
j}}^N\frac{u+a_\ell}{a_j-a_\ell}.
\end{equation}
Clearly, the sum can be written as the addition of all the residues
at $v=a_\ell$ ($\ell=1,\ldots,N$) of
\[\,_0F_1(\alpha+1,vy)e^{-v}\prod_{\ell =1}^N\frac{u+a_\ell}{v-a_\ell}\frac{1}{u+v}.\]
This proves Eq.\ \eqref{Kfirst} when all $a_j$'s are distinct.  The validity of this expression for the general case is established by continuity.  The quadruple integral representation of the kernel is readily obtained by comparing Eqs.\ \eqref{BesselI} and \eqref{I0F1}.\end{proof}

Multiple functions associated to the modified Bessel function of the
first kind have been recently introduced by Coussement and Van
Assche \cite{CVA1}.  Their work, which was not motivated by Random
Matrix Theory, corresponds to our $D=1$ case. Here we provide
integral representations of the type I and II multiple functions for
$D\geq 1$.

To avoid any confusion between the number of weights and the number
of distinct eigenvalues in the perturbation matrix, we have to
introduce some new notations.  Suppose that
$\mathbf{a}=\mathbf{b}^{\vec{m}}$, where $\vec{m}=(m_1,\ldots,m_d)$,
$|\vec{m}|=N$, and $b_1>b_2>\ldots>b_d> 0$. Set
$\vec{n}=(n_1,\ldots,n_D)$ with
\begin{align}
n_1&=\left\lfloor\frac{m_1+1}{2}\right\rfloor,&
n_2&=\left\lfloor\frac{m_1}{2}\right\rfloor,\nonumber\\
n_3&=\left\lfloor\frac{m_2+1}{2}\right\rfloor,&
n_4&=\left\lfloor\frac{m_2}{2}\right\rfloor,\nonumber\\
&\vdots&&\vdots&\nonumber\\
n_{D-1}&=\left\lfloor\frac{m_d+1}{2}\right\rfloor&n_D&=\left\lfloor\frac{m_d}{2}\right\rfloor
\end{align} where $D=2d$.  If $b_d=0$, it is understood that $n_D=m_d$ and $D=2d-1$.
The multi-index $\vec{n}$ gives the correct multiplicities of the
weight functions (see Section 3); that is, if $b_d>0$,
\begin{equation}
\vec{w}(x)=\Big[w_\alpha(x,b_1),\,w_{\alpha+1}(x,b_1),\,w_\alpha(x,b_2),\,w_{\alpha+1}(x,b_2),\,\ldots,\,w_\alpha(x,b_d),\,w_{\alpha+1}(x,b_d)\Big]
\end{equation}
or, if $b_d=0$,
\begin{equation}
\vec{w}(x)=
\Big[w_\alpha(x,b_1),\,w_{\alpha+1}(x,b_1),\,w_\alpha(x,b_2),\,w_{\alpha+1}(x,b_2),\,\ldots,\,w_\alpha(x,b_d)\Big],
\end{equation}
where $w_\alpha$ stands for the function defined in Eq.\
\eqref{walpha} with $V(x)=x$ .

\begin{proposition} Following the above notation,  we
have that the multiple function of type I is
\begin{align}\label{intQ}
Q_{\vec{n}}(x)&=w_\alpha(x)\int_{\mathcal{C}_{\mathbf{a}}}\frac{dv}{2\pi\im}\frac {e^{-v}\,_0F_1(\alpha+1,xv)}{\prod_{i=1}^N(v-a_i)}\\
&=\Gamma(\alpha+1)w_\alpha(x)\int_{\mathcal{C}_{\mathbf{a}}}\frac{dv}{2\pi\im}
\int_{\mathcal{C}_{\{0\}}}\frac{dz}{2\pi\im}
\frac{z^{\alpha-1}e^{v(xz-1)+1/z}}{\prod_{i=1}^N(v-a_i)},
\end{align}
while the multiple polynomial of type II is given by
\begin{align}\label{intP}
P_{\vec{n}}(x)&=\frac{(-1)^{N}e^x}{\Gamma(\alpha+1)}\int_0^\infty du\,u^\alpha e^{-u}\,_0 F_1(\alpha+1,-xu)\prod_{i=1}^N(u+a_i)\\
&=\frac{(-1)^{N}}{\Gamma(\alpha+1)w_{\alpha}(x)}\int_0^\infty
du\,\int_{\mathcal{C}_{\{0\}}}\frac{dw}{2\pi\im}
\frac{e^{-u(xw+1)+1/w}}{w^{\alpha+1}} \prod_{i=1}^N(u+a_i).
\end{align}
\end{proposition}
\begin{proof}We start with the determinant representation of the
type  I function given in Proposition 2.  Here,
\[Z_N=Z'_N\Delta(a_1,\ldots,a_N)=N!\prod_{i=1}^Ne^{a_i}\Delta(a_1,\ldots,a_N).\]
Eq.\ \eqref{detQ} then implies
\[Q_{\vec{n}}(x)=\sum_{i=1}^N\frac{\xi_i(x)e^{-a_i}}{\prod_{j\neq i}(a_i-a_j)}
=w_\alpha(x)\sum_{i=1}^N\frac{e^{-a_i}\,_0F_1(\alpha+1,a_ix)}{\prod_{j\neq
i}(a_i-a_j)},\] which turns out to be equivalent to the proposed
formulae.

The proof for the type II function is a bit more tricky.  Firstly,
Eq.\eqref{detP} gives
\[P_{\vec{n}}(x)=\frac{(-1)^N}{\Delta(a_1,\ldots,a_N)}
\sum_{n=1}^N(-1)^{n}\eta_n(x)\left|\begin{array}{c}\left[a_j^{i-1}\right]_{\substack{i=1,\ldots,n-1\\j=1,\ldots,N}}\\
\left[a_j^{i-1}\right]_{\substack{i=n+1,\ldots,N\\j=1,\ldots,N}}\end{array}\right|.
\]Secondly,
we introduce the elementary symmetric functions, denoted by $e_n$,
and defined via the following generating function:
\[\prod_{i=1}^N(t+a_i)=:\sum_{n=0}^Nt^ne_{N-n}(a_1,\ldots,a_N).\]
Then a few manipulations give
\begin{equation*}
\frac{1}{\Delta(a_1,\ldots,a_N)}
\left|\begin{array}{c}\left[a_j^{i-1}\right]_{\substack{i=1,\ldots,n-1\\j=1,\ldots,N}}\\
\left[a_j^{i-1}\right]_{\substack{i=n+1,\ldots,N\\j=1,\ldots,N}}\end{array}\right|=e_{N+1-n}(a_1,\ldots,a_N)
\end{equation*}Hence
\begin{equation}\label{PLaguerre}
P_{\vec{n}}(x)=(-1)^N\sum_{n=0}^N(-1)^n\eta_{n+1}(x)e_{N-n}(a_1,\ldots,a_N).
\end{equation}
 Thirdly, we exploit the integral representation \eqref{L0F1} to obtain
\[P_{\vec{n}}(x)=\frac{(-1)^Ne^x}{\Gamma(\alpha+1)}
\sum_{n=0}^N\int_0^\infty dt\, t^\alpha
e^{-t}\,_0F_1(\alpha+1,-xt)\, t^n e_{N-n}(a_1,\ldots,a_N).\] The
first integral representation of $P_{\vec{n}}$ is finally obtained
by reconstructing the generating function of the elementary
symmetric functions. The last double integral expression is a mere
consequence of Eqs.\ \eqref{BesselI} and \eqref{I0F1}.
\end{proof}

 The above multiple functions can be considered as multi-parameter
generalizations of the Laguerre polynomials.  Indeed, from the
integral representations of the Laguerre polynomials,
\begin{equation}L^{\alpha}_N(x)
=\int_{\mathcal{C}_{\{0\}}}\frac{dw}{2\pi\mathrm{i}}\frac{e^{-xw}}{w^{N+1}}(1+w)^{N+\alpha}
=\frac{(N+\alpha)!}{N!}x^{-\alpha}
\int_{\mathcal{C}_{\{0\}}}\frac{dw}{2\pi\mathrm{i}}\frac{e^{xw}}{w^{N+\alpha+1}}(w-1)^{N}
\end{equation} when $\alpha\in\mathbb{Z}$, one can show that
\begin{equation}
\lim_{\mathbf{a}\rightarrow
\mathbf{0}}Q_{\vec{n}}(x)=\frac{(-1)^{|\vec{n}|-1}}{(|\vec{n}|+\alpha-1)!}\,x^{\alpha}e^{-x}L^{\alpha}_{|\vec{n}|-1}(x)
\end{equation}and
\begin{equation}
\lim_{\mathbf{a}\rightarrow
\mathbf{0}}P_{\vec{n}}(x)=(-1)^{|\vec{n}|}|\vec{n}|!\,L^{\alpha}_{|\vec{n}|}(x).
\end{equation}
Note that the last identity is more easily shown by using Eq.\
\eqref{PLaguerre}, i.e., \begin{equation}
P_{\vec{n}}(x)=(-1)^N\sum_{n=0}^Nn!\,e_{N-n}(a_1,\ldots,a_N)\,L^{\alpha}_n(x),
\end{equation}
which is valid for all $a_1,\ldots,a_N$.  Note that the formulae
\eqref{intQ} and \eqref{intP} furnish an alternative method to
prove some of the properties  found in \cite{CVA1} for the
multiple polynomials with $D=1$.

\begin{corollary}
Let $a_1>a_2>\ldots>a_N\geq 0$. Let also $n^{(0)}=(0,0,\ldots,0)$,
$n^{(1)}=(1,0,0,\ldots,0)$, $n^{(2)}=(1,1,0,\ldots,0)$, and so on
till  $n^{(N)}=(1,1,\ldots,1)$.  Define $P_i=P_{\vec{n}^{(i)}}$ and
$Q_i=Q_{\vec{n}^{(i+1)}}$.  Then
\begin{equation}
K_N(x,y)=\sum_{i=0}^{N-1}P_i(x)\,Q_i(y).
\end{equation}
\end{corollary}
\begin{proof}
We substitute the formula \cite{DF0406}
\[\frac{1}{u-v}\prod_{i=1}^N\frac{u+a_i}{v+a_i}=\frac{1}{u-v}+
\sum_{k=1}^N\frac{\prod_{i=1}^{k-1}(u+a_i)}{\prod_{i=1}^{k}(v+a_i)}\]
in \eqref{Kfirst}. We then  use the fact that
\[\int_{\mathcal{C}^{\{u\}}_{-\mathbf{a}}}\!\!\frac{dv}{2\pi\im}
\frac{e^v\,_0F_1(\alpha+1,-yv)}{u-v}=0\] and obtain
\[K_N(x,y)=\sum_{k=1}^N\frac{y^\alpha
e^{-y+x}}{\Gamma(\alpha+1)^2}\!\int_0^\infty du\,u^\alpha e^{-u}\,_0
F_1(\alpha+1,-xu)\prod_{i=1}^{k-1}(u+a_i)\!\!\int_{\mathcal{C}^{\{u\}}_{-\mathbf{a}}}\!\!\frac{dv}{2\pi\im}
\frac{e^v\,_0F_1(\alpha+1,-yv)}{\prod_{j=1}^{k}(v+a_j)}.\] The
comparison with Eqs.\ \eqref{intQ} and \eqref{intP} finishes the
proof.
\end{proof}

 The general Christoffel-Darboux (CD) formula involving multiple polynomials
 of type I and II has been found by Daems and Kuijlaars \cite{DK}.
 The complicated relation between the weights and the perturbation
 eigenvalues makes difficult the extraction of the general CD formula
 from the integral representation of the kernel.  This is in
 contradistinction with the perturbed
 Laguerre ensemble in which the CD formula is readily derived from
 integration by parts \cite{BKint, DF0406}

\section{Concluding remarks}

 The integral representations of both the
kernel and multiple functions of type I and II provide tools for
studying the asymptotic behavior of the perturbed chGUE. Of
particular interest are the finite rank perturbations
\cite{Baik2004, Baik2005,Peche}. In such systems, the eigenvalues of
the perturbation matrix $\mathbf{A}$ satisfy
\[ \mathbf{a}=(a_1,\ldots,a_r,0,\ldots,0).\]
The rank of the perturbation is given by $r$; it is finite in the
sense that $\lim_{N\rightarrow\infty}N^{-1}r=0$. It has been
observed in \cite{DF0406} that the kernel of the perturbed Laguerre
ensemble can be decomposed as a unperturbed kernel $\bar{K}_{N-r}$,
where
\[\bar{K}_{N}(x,y)=
\frac{N!}{(N+\alpha-1)!}\frac{y^{\alpha} e^{-y}}{x-y}\Big(
L^\alpha_{N-1}(x)L^\alpha_{N}(y)-L^\alpha_{N}(x)L^\alpha_{N-1}(y)\Big),\]
 plus a sum of $r$ projectors $p_i\otimes q_i$, where $p_i$ and
$q_i$ are respectively related to the type II and I multiple
Laguerre functions. A similar decomposition  exists in the perturbed
chGUE; explicitly,
\[K_N(x,y)=\bar{K}_{N-r}(x,y)+\sum_{k=1}^rp_k(x)\,q_k(y), \]
where
\begin{align*}
p_k(x)&=\frac{e^x}{\Gamma(\alpha+1)}\int_0^\infty
du\,u^{N+\alpha-r}\prod_{i=1}^{k-1}(u+a_i) e^{-u}\,_0
F_1(\alpha+1,-xu),\\
q_k(x)&=\frac{x^\alpha
e^{-x}}{\Gamma(\alpha+1)}\int_{\mathcal{C}_{-\mathbf{a}}}\frac{dv}{2\pi\im}\frac
{e^{v}\,_0F_1(\alpha+1,-xv)}{v^{N-r}\prod_{i=1}^k(v+a_i)}.
\end{align*} The asymptotic correlations of the chGUE with perturbation will be considered
in a forthcoming paper.

\begin{acknow}
The work of P.J.F. has been supported by the Australian Research Council.
P.D. is grateful to the Natural Sciences and Engineering Research Council
of Canada for a postdoctoral  fellowship. We thank E.~Kanzieper for drawing 
our attention to the question of seeking a general formula such as
(\ref{Q}) for $Q(x)$.
\end{acknow}

\end{document}